\documentclass[12pt]{iopart}
\usepackage{iopams}
\usepackage{graphicx}
\newcommand{\beq}{\begin{equation}}
\newcommand{\eeq}{\end{equation}}

\newcommand{\beqa}{\begin{eqnarray}}
\newcommand{\eeqa}{\end{eqnarray}}
\newcommand{\beqas}{\begin{eqnarray*}}
\newcommand{\eeqas}{\end{eqnarray*}}

\usepackage{tikz}
\usetikzlibrary{positioning}

\begin{document}
\title{Transient Non-Confining Potentials for Speeding Up a Single Ion Heat Pump}

\author{E. Torrontegui$^{1}$, S. T. Dawkins$^{2}$, M. G{\"o}b$^{2}$, K. Singer$^{2}$}

\address{$^{1}$ Instituto de F\'{\i}sica Fundamental IFF-CSIC, Calle Serrano 113b, 28006 Madrid, Spain}
\address{$^{2}$ Experimentalphysik I, Universit\"at Kassel, Heinrich-Plett-Str. 40, 34132 Kassel, Germany}
\ead{ \mailto{eriktorrontegui@gmail.com}, \mailto{ks@uni-kassel.de}}

\begin{abstract}
We propose speeding up a single ion heat pump based on a tapered ion trap. If a trapped ion is excited in an oscillatory motion axially the radial degrees of freedom are cyclically expanded and compressed such that heat can be pumped between two reservoirs coupled to the ion at the turning points of oscillation. Through the use of invariant-based inverse engineering we can speed up the process without sacrificing the efficiency of each heat pump cycle.  This additional control can be supplied with additional control electrodes or it can be encoded into the geometry of the radial trapping electrodes. 
We present novel insight how speed up can be achieved through the use of inverted harmonic potentials and verified the stability of such trapping conditions.
\end{abstract}
%
%
%
%
%
%
\section{Introduction}
\begin{figure}[htbp]
\begin{center}
\includegraphics[width=10cm]{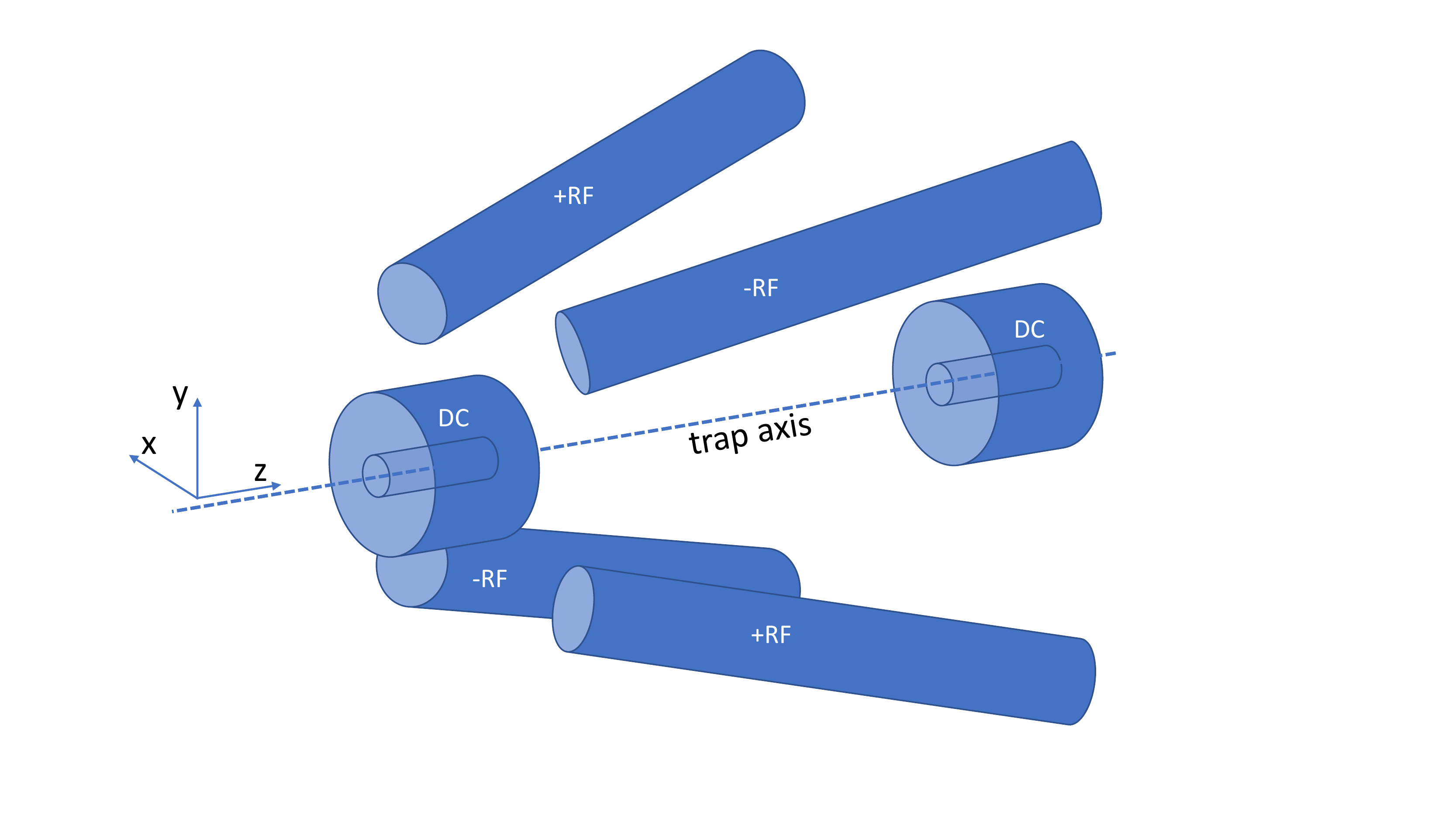}
\caption{Tapered ion trap. The tapered electrodes are supplied symmetrically with radiofrequency voltage for the radial confinement. Endcaps are used to supply axial confinement with dc voltages. During shortcut to adiabaticity protocol the radiofrequency is switched off and the voltage on the endcaps is used to realize the axial confinement or anti-confinement respectively. Due to the short duration stable trapping conditions can be maintained.}
\label{fig:tapered trap}
\end{center}
\end{figure}

Trapped ions are an established platform for realizing high-fidelity quantum information processing~\cite{blatt2008entangled,schafer2018fast}, quantum simulation~\cite{friedenauer2008simulating,zhang2018experimental}, and precision metrology experiments~\cite{rosenband2008frequency,schulte2016quantum}.
Recently a single ion, trapped in a tapered trap, was employed to realize a single ion heat engine~\cite{rossnagel2016single,abah2012single}.
Due to the controllability of the environment this system implements a formidable model experiment for studying thermodynamics at the single particle limit towards the quantum regime.
In this paper we study the reverse process, a single ion heat pump, and how this process can be sped up through the shortcut to adiabaticity technique involving the use of invariant-based inverse engineering
\cite{Chen2010_063002, Torrontegui2013_117}.
In the following, as in the single ion heat engine, the ion is confined in a harmonic potential and the motional radial degrees of freedom serve as the working agent, where we consider temperature only in the radial directions.
A thermal state that is adiabatically transported along the taper (see Fig.~\ref{fig:tapered trap}) into a region with lower trap frequency attains a lower temperature due to the reduced energy level spacing in the harmonic potential.
This mechanism could be used to couple to a reservoir, such as neighbouring ions, to affect cooling by absorbing heat.  
Thus, a subsequent adiabatic transport back to the starting position at higher confinement results in an increased temperature.
Dumping heat to another reservoir allows one to recool the working agent for starting a new cycle of the heat pump.
Speeding this procedure up to increase the heat pumping rate through the use of e.g. bang-bang transport is typically limited by the condition of performing the change of the radial trapping frequency adiabatically.
In the following, we will describe how shortcuts to adiabaticity can be employed to go beyond this limitation \cite{DEng2013, del2014more, Beau2016, Chotorlishvili2016, Deng2018}, in particular
the invariant-based inverse engineering approach will allow the design of protocols by controlling the radial trapping frequency with external electrode voltages.
One possibility is to control the radial trapping frequency by varying the radiofrequency amplitude which is symmetrically supplied to the tapered electrodes of the ion trap.
The speed up in this case is limited by the fact that the radial confinement should be sustained. 
A further speed up would be possible if the trapping potential can be inverted.
This is achieved by switching off the radiofrequency confinement for a short period and using the radial DC potentials generated by the end-cap electrodes to supply a specially designed time varying radial quadratic potential.
Due to Laplace's equation, a confining potential in one direction leads necessarily to repelling potentials in the other two directions, or vice versa.
We present a shortcut of short duration, which helps both to achieve high cooling rates and avoid losing the ion from the trap.
For typical trapping frequency changes from 3\,MHz to 1\,MHz, a shortcut duration of 20\,ns can be achieved, through the use of non-confining potentials.
In order to avoid instability due to micromotion, it is necessary that the radiofrequency period is shorter than the shortcut duration. \footnote{Note that the switching should be synchronized to the radiofrequency phase}
Numerical simulations confirm stable trapping conditions despite the inverted trapping potentials over short time periods. It is important to note that the speed up is only limited by the maximal voltages and the currents which can be applied to the electrodes.
The single ion heat pump could be an important method for lowering temperatures in a trapped ion based quantum information processor and the speed up described could help to compete against deleterious heating rates. 

\section{Invariant-based inverse engineering for mixed states}
Closed quantum systems follow a unitary dynamics described by the Liouville equation of motion
\beq
\label{Lio}
i\hbar\frac{\partial\hat\rho(t)}{\partial t}=[\hat H(t),\hat\rho(t)]
\eeq
where $\hat\rho(t)$ is the density matrix describing the system and $\hat H(t)$ the Hamiltonian  controlling its dynamics.
Related to any Hamiltonian there are dynamical invariants of motion \cite{Lewis1969}
\beq
\label{dynI}
i\hbar\frac{\partial\hat I(t)}{\partial t}-[\hat H(t),\hat I(t))]=0,
\eeq
with constant expectation values, --i.e. quantities preserved by the dynamics generated by (\ref{Lio}). The invariant
expands an orthonormal basis $|\phi_n(t)\rangle$ with constant eigenvalues $\lambda_n$,
\beq
\label{inv}
\hat I(t)=\sum_n|\phi_n(t)\rangle\lambda_n\langle\phi_n(t)|.
\eeq
In this basis the density matrix elements $\rho_{lk}\equiv\langle\phi_l(t)|\hat\rho(t)|\phi_k(t)\rangle$ are calculated from \cite{Levy2018}
\beqa
\label{LR_rho}
\dot\rho_{lk}(t)&=&i\bigg(\langle\phi_l(t)\bigg| i\hbar\frac{\partial}{\partial t}-\hat H(t)|\bigg|\phi_l(t)\rangle-\langle\phi_k(t)\bigg| i\hbar\frac{\partial}{\partial t}-\hat H(t)|\bigg|\phi_k(t)\rangle\bigg)\rho_{lk}(t) \nonumber\\
\dot\rho_{kk}(t)&=&0
\eeqa
where the populations remain constant and the off-diagonal elements depend on the difference of time derivatives of two Lewis-Riesenfeld phases \cite{Lewis1969}.
A simpler derivation of the Lewis-Riesenfeld relation for pure states is done in \ref{LR_pure}. From Eq. (\ref{LR_rho}) we observe that a system initialized in an eigenstate of the invariant
will remain in the same instantaneous eigenstate without transitions, imposing the so-called frictionless conditions $[\hat H(0),\hat I(0)]=[\hat H(t_f),\hat I(t_f)]=0$, we ensure that the
system starts and ends as an eigenstate of the Hamiltonian without unwanted excitations. A perfect state transfer from $\hat H(0)$ to $\hat H(t_f)$ is designed by first choosing
properly $\hat I(t)$ and then reverse engineering the dynamics to deduce $\hat H(t)$.
In particular, for an effectively 1D time dependent harmonic potential
\beq
\label{H1d}
\hat H(t)=\frac{\hat p^2}{2m}+\frac{1}{2}m\omega^2(t)\hat q^2
\eeq
an associated dynamical invariant (\ref{dynI}) reads \cite{Lewis1982}
\beq
\label{iho}
\hat I(t)=\frac{1}{2m}[b(t)p-m\dot b(t)q]^2+\frac{1}{2}m\omega_0^2\frac{q^2}{b^2(t)},
\eeq
where $b(t)$ is a free function of time satisfying the Ermakov equation \cite{Ermakov1880}
\beq
\label{erm}
\ddot b(t)+\omega^2(t)b(t)=\frac{\omega_0^2}{b^3(t)},
\eeq
being $\omega_0$ the initial frequency of the oscillator at time $t=0$. The frictionless conditions $[\hat H (t_b)=\hat I(t_b)]=0$ at the boundary times $t_b=0, t_f$ set
\beqa
\label{BC}
b(0)&=&1,\quad\dot b(0)=0,\quad\ddot b(0)=0\nonumber\\
b(t_f)&=&\gamma,\quad\dot b(t_f)=0,\quad\ddot b(t_f)=0
\eeqa
with $\gamma=(\omega_0/\omega_f)^{1/2}$. Any $b(t)$ fulfilling the previous six boundary conditions will produce a perfect control, see Eq. (\ref{erm})
\beq
\label{w}
\omega^2(t)=\frac{\omega_0^2}{b^4(t)}-\frac{\ddot b(t)}{b(t)}
\eeq
driving each
Fock state $|n(0)\rangle$, to the corresponding Fock state $|n(t_f)\rangle$ independently of the process time $t_f$.
More details in \ref{reverse} can be found.
Note that typically for ultrafast processes, very short $t_f$ values, $\omega_0^2/b^4<\ddot b/b$ and the trapping parabola becomes a repeller potential. The stability and experimental
implementation of such scenario will be deeply analyzed in the following sections.
\section{Reverse engineering of Gaussian states}
\subsection{Coherent states}
The previous protocol (\ref{w}) is not only valid to connect single $|n\rangle$ to $|n\rangle$ Fock states but also coherent states \cite{Glauber1963}
\beq
|\alpha(t)\rangle=e^{-|\alpha|^2/2}\sum_{n=0}^\infty\frac{\alpha^n}{\sqrt{n!}}|n(t)\rangle.
\eeq
These are pure states forming a linear superposition. As at initial time the frictionless conditions guarantee that $\hat H$ and $\hat I$ share a common basis $|\phi_n(0)\rangle=|n(0)\rangle$ and according to Eq. (\ref{LR_rho}),
or simply (\ref{wf}) as the system is pure, this initial state
$|\psi(0)\rangle=|\alpha(0)\rangle$ will evolve to \cite{Palmero2016}
\beq
\label{coh_state}
|\psi(t_f)\rangle=e^{-ig\omega_0/2}e^{-|\tilde\alpha|^2/2}\sum_{n=0}^{\infty}\frac{\tilde\alpha^n}{\sqrt{n!}}|\phi_n(t_f)\rangle=|\tilde\alpha(t_f)\rangle,
\eeq
with $\tilde\alpha=\alpha e^{-ig\omega_0}$ and $g=\int_0^{t_f}dt'/\rho^2$. The condition $[\hat H(t_f),\hat I(t_f)]=0$ guarantees $|\phi_n(t_f)\rangle=|n(t_f)\rangle$, thus the system ends as also a coherent state with frequency $\omega_f$.
\subsection{Thermal states}
From the set of Eqs. (\ref{LR_rho}) we observe that any system that initially is diagonal in the basis expanded
by the eigenstates of the invariant will keep its populations constant during the whole process. Moreover, imposing $[\hat I (t_b),\hat H(t_b)]=0$ at $t_b=0,t_f$ the initial and final states will be also diagonal in the energy basis
expanded by $\hat H(0)$ and $\hat H(t_f)$, which is the case for thermal states.
Considering the time-dependent harmonic oscillator (\ref{H1d}) and if initially the system is assumed to be the thermal state $\hat\rho(0)=\exp(-\beta_0\hat H(0))/Z$, with $Z$ a normalization constant, initial inverse temperature
$\beta_0$, and $\omega(t=0)=\omega_0$, by changing $\omega(t)$ according to Eqs. (\ref{w}) and (\ref{BC}) the system will evolve reaching the final thermal state $\hat\rho(t_f)=\exp(-\beta_f\hat H(t_f))/Z^{\prime}$, 
corresponding to a $\hat H(t_f)$ with frequency $\omega(t_f)=\omega_f$ and a cooling/heating $\beta_f=\gamma^2\beta_0$.
\subsection{Quantum dynamical evolution of Gaussian states}
Note that both coherent and thermal states are  Gaussian states, --i.e. the symmetric Wigner function
\beq
W(\mathbf{x})=W(q, p)=\frac{1}{\pi\hbar}\int_{-\infty}^{\infty}dy\langle q+y|\hat\rho|q-y\rangle e^{-2ipy/\hbar}
\eeq
is Gaussian, $\mathbf{x}\equiv (q, p)$ corresponds to the eigenvalues of the quadrature operators $\hat\mathbf{x}\equiv (\hat q, \hat p)$. Consequently, the density operator $\hat{\rho}$ has a one-to-one correspondence
with the first and second-order statistical moments of the state, $\hat\rho\equiv\hat\rho(\bar\mathbf{x},\mathbf{V})$ \cite{Weedbrook2012}. The first moments are called the displacement vector, or simply the mean value
\beq
\bar\mathbf{x}=\langle\hat x_i\rangle=\Tr [\hat x_i\hat\rho(t)],
\eeq
and the second moment, called covariant matrix, with generic element
\beq
\mathbf{V}=V_{ij}=\frac{1}{2}\langle\{\Delta\hat x_i,\Delta\hat x_j\}\rangle,
\eeq
where $\Delta\hat x_i=\hat x_i-\langle\hat x_i \rangle$ and $\{\hat A,\hat B\}=\hat A\hat B+\hat B\hat A$. In particular, for coherent and thermal states of a harmonic oscillator these moments $\bar\mathbf{x}$ and
$\mathbf{V}$ are constructed from the set of operators $\hat\mathbf{X}\equiv (\hat q,\hat p, \hat q^2,\hat p^2, \hat q\hat p+\hat p\hat q)$,
\beq
\label{moments}
\bar\mathbf{x}=(\langle\hat q\rangle,\langle\hat p\rangle),\quad \mathbf{V}=\bigg(
\begin{array}{cc}
\langle\hat q^2\rangle-\langle\hat q\rangle^2 & \langle\hat q\hat p+\hat p\hat q \rangle-\langle \hat p\rangle\langle \hat q\rangle \nonumber\\
\langle\hat q\hat p+\hat p\hat q \rangle-\langle \hat p\rangle\langle \hat q\rangle & \langle\hat p^2\rangle-\langle\hat p\rangle^2
\end{array}
\bigg)
\eeq
and the Wigner function is reconstructed,
\beq
\label{wigner}
W(\mathbf{x})=\frac{\exp[(\mathbf{x}-\bar\mathbf{x})^T\mathbf{V}^{-1}(\mathbf{x}-\bar\mathbf{x})/2]}{2\pi\sqrt{\det \mathbf{V}}}
\eeq
 with $\mathbf{x}^T$, the transpose of $\mathbf{x}$ and $\mathbf{V}^{-1}$ the inverse matrix of $\mathbf{V}$.
In order to describe the dynamical evolution of $\hat \rho$, or equivalently $W(\mathbf{x})$, it is enough to describe the evolution of the set of observables $\hat\mathbf{X}$
to reconstruct the state using Eqs. (\ref{moments}) and (\ref{wigner}), avoiding the use
of wave packet propagation. This is done within the Heissenberg representation
\beq
\label{motion}
\frac{d\bar{\mathbf{X}}_i(t)}{dt}=\frac{i}{\hbar}[\hat H(t),\bar{\mathbf{X}}_i(t)],
\eeq
with $\bar{\mathbf{X}}_i(t)\equiv \langle \hat{\mathbf{X}}_i\rangle=\Tr [\hat{\mathbf{X}}_i\hat\rho(t)]$.
Note that the set of five operators $\hat{\mathbf{X}}$ form a closed Lie algebra, as the Hamiltonian (\ref{H1d}) of a harmonic oscillator is a linear combination of some $\hat{\mathbf{X}}_i$ elements, the dynamical
equation of motion (\ref{motion}) is also closed to the algebra. Consequently, the evolved state $\hat\rho(t)$ remains Gaussian during the whole evolution.

Finally, given two Gaussian states $\hat\rho_1$ and $\hat\rho_2$,
we can compute the fidelity $\mathcal{F}(\hat\rho_1,\hat\rho_2)=\Tr(\sqrt{\sqrt{\hat\rho_1}\hat\rho_2\sqrt{\hat\rho_1}})$
between these two states in terms of their respective moments $\bar{\mathbf x}_1, \mathbf{V}_1$ and $\bar{\mathbf x}_2, \mathbf{V}_2$ as
\beq
\mathcal{F}(\hat\rho_1,\hat\rho_2)=\mathcal{F}_0(\hat V_1,\hat V_2)\exp\bigg[-\frac{1}{4}\delta_{\bar{\mathbf{x}}}^T(\mathbf{V}_1+\mathbf{V}_2)^{-1}\delta_{\bar{\mathbf{x}}}\bigg]
\eeq
with $\delta_{\bar{\mathbf{x}}}=\bar{\mathbf{x}}_2-\bar{\mathbf{x}}_1$ and $\mathcal{F}_0(\hat V_1,\hat V_2)$ having a closed analytical form \cite{Banchi2015}.
\section{Robustness improvements}
The main source of imperfection in the experimental implementation of the shortcut is produced by the time variation of the control $\omega^2(t)$. Controlling this by the pseudopotential through dynamic change of the amplitude of the radio-frequency voltage has the disadvantage that non-confining potentials cannot be supplied. Amplitude control of this voltage is technologically more involved and intrinsically limited by the period of the radiofrequency. Thus the biggest speed up potential and controllability is obtained by controlling the DC potentials by low-noise high-speed arbitrary waveform generators. If radiofrequency confinement is kept on very accurate timing and high voltages are needed. In order to allow for a reliable control of the confinement, we therefore switch off the radiofrequency drive during the control period. This can be efficiently achieved by a solid state radiofrequency toggle switch \cite{4743596} directly after a high voltage rf generator \cite{doi:10.1063/1.1148297}. In many cases the high voltage rf generator is replaced by a low voltage radiofrequency generator with a subsequent radiofrequency amplifier with 50$\Omega$ impedance. Impedance matching is then achieved with a helical responators which additionally transforms the radiofrequency voltages. In these cases an ultra low resistance toggle switch has to be used directly after the helical resonator with one terminal connected with the trap electrodes and the other connected with a circuit of equivalent impendance. Anharmonicities of the trapping potentials can be neglected as the ion is kept at the extremal point of the harmonic confinement at all the time.

Thanks to the freedom in the construction of the shortcut protocol at intermediate time more constraints such as minimizing $d\omega^2/dt\equiv\partial_t(\omega^2)$ due to experimental limits can be realized. This is originated from the slew rate and bandwidth limit of digital analog converters and power amplifiers. The minimization of $\partial_t(\omega^2)$ can then be performed by optimal control techniques but the boundary conditions for $b$ could violated. Discontinuities in $\dot b, \ddot b$ would be unfeasible due to the requirement of instantaneous jumps in the control voltages.

\begin{figure}[t]
\begin{center}
\includegraphics[width=0.5 \textwidth]{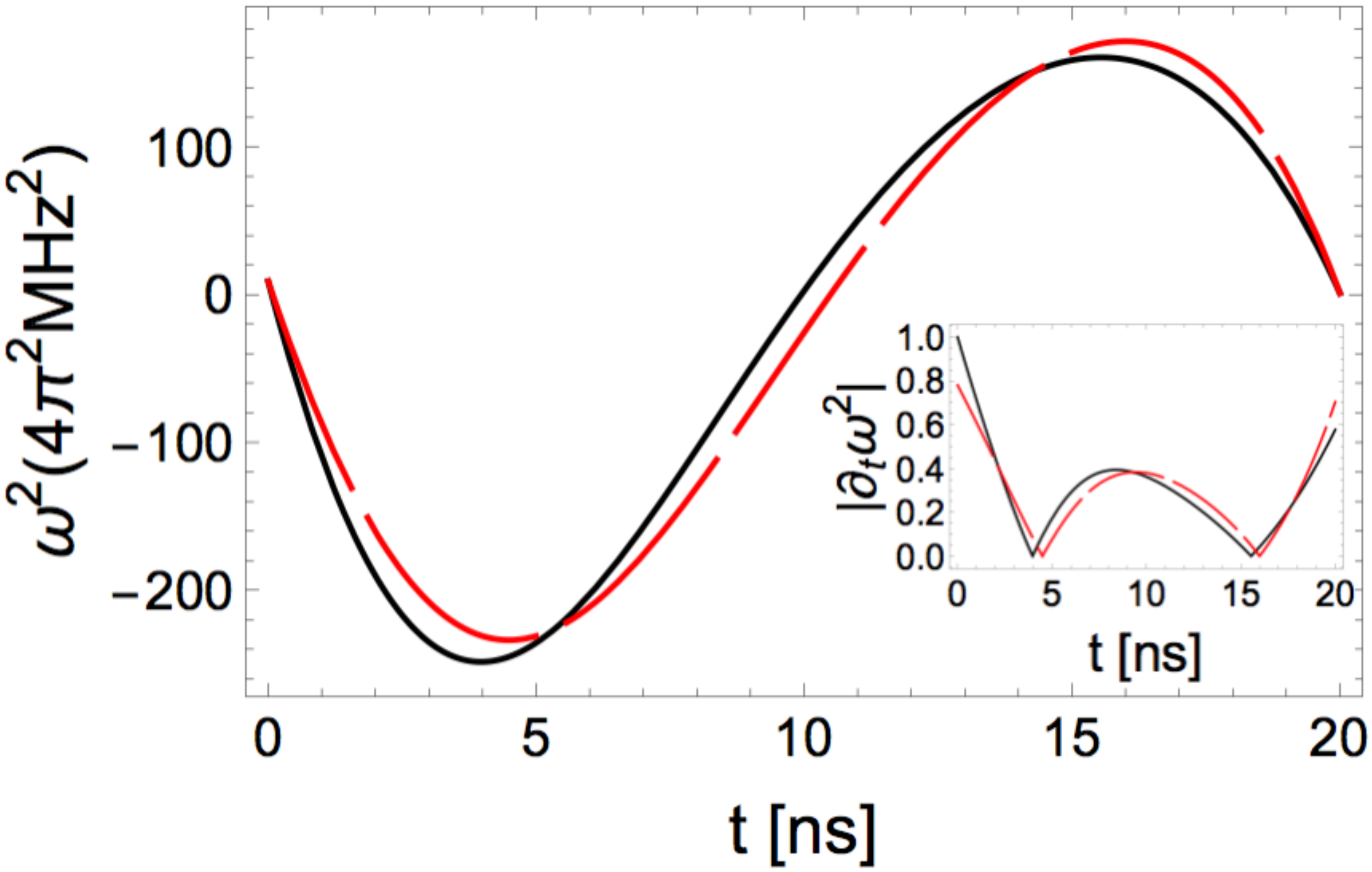}
\caption{Radial confinement $\omega^2$ as a function of time for two different designs of the shortcut. (Black-solid line) standard 6-order polynomial fulfilling the  frictionless conditions (\ref{BC}). (Red long-dashed line) Improved design with extra-coefficients to control
the value of $\max|\partial_t(\omega^2)|$. The inset shows $|\partial_t\omega^2(t)|/|\partial_t\omega^2(0)|$ for both designs.
Here, $\omega_{\mathrm{0}}/(2 \pi)$=3\,MHz and $\omega_{\mathrm{f}}/(2 \pi)$=1\,MHz. }
\label{fig_omegas}
\end{center}
\end{figure}

As an example, minimizing $\max|\partial_t(\omega^2)|$, the maximum value of $\partial_t(\omega^2)$ in the interval $t\in [0,t_f]$, will reduce the power employed by the control protocol improving the heat extraction process. Defining $\mathcal{C}(t)=\omega^2(t)$,
the extreme condition that minimizes $d\mathcal{C}/dt=0$ is satisfied by the useless control $\mathcal{C}(t)=\omega^2_{opt}(t)=const$. The mean value theorem provides a useful bound
for the instantaneous maximum value of the control. Assuming that $\mathcal{C}$ is continuous in $[0,t_f]$ and differentiable in $(0,t_f)$ such that $\mathcal{C}(0)=\omega_0^2$ and $\mathcal{C}(t_f)=\omega_f^2$
the maximum of its derivative must be
\beq
\frac{d\omega^2}{dt}\ge\frac{\omega^2_0-\omega_f^2}{t_f}
\eeq
where the equality holds for the $\omega^2(t)=\omega_0^2+(\omega_f^2-\omega_0^2)t/t_f$ control. However, the resulting $b(t)$ deduced from Eq. (\ref{erm}) does not satisfy the six frictionless boundary conditions (\ref{BC}). As result
discontinuities in $\dot b$ and $\ddot b$ at $t=0$ and $t_f$ should be applied requiring instantaneous switches in the controls.
In order to avoid discontinuities hardly resolved experimentally we use the non-uniqueness of $b(t)$
to add extra-parameters $a_i $ in the interpolation of $b(t)=\sum_i a_it^i$ to ensure (\ref{BC}) and using Eq. (\ref{w}) create controls $\omega^2(t;a_i)$ such that the value of $\partial_t(\omega^2)$ is controlled through thee extra-parameters $a_i$ \cite{Levy2018, Levy2017}.
By using gradient descent methods $\omega^2(t;a_i)$ is optimized. As an example, for an expansion process of $20$ ns see Fig. \ref{fig_omegas},
the addition of the extra-coefficient $a_6t^6$ in the interpolation of $b(t)$ allows a reduction of $\frac{\max|\partial_t(\omega^2_{opt})|}{\max|\partial_t(\omega^2)|} \sim 0.78$ in contrast with
a standard 6 order interpolation, see \ref{reverse}. Additionally, this design also reduces the value of $\max|\omega^2|$, thus the protocol improves both the slew rate and power of the required controls.
Other sophisticated designs are also possible due
to the freedom to interpolate $b(t)$ at intermediate times.
\section{Proposed experimental implementation}
\begin{figure}[t]
\begin{center}
\includegraphics[width=0.45 \textwidth]{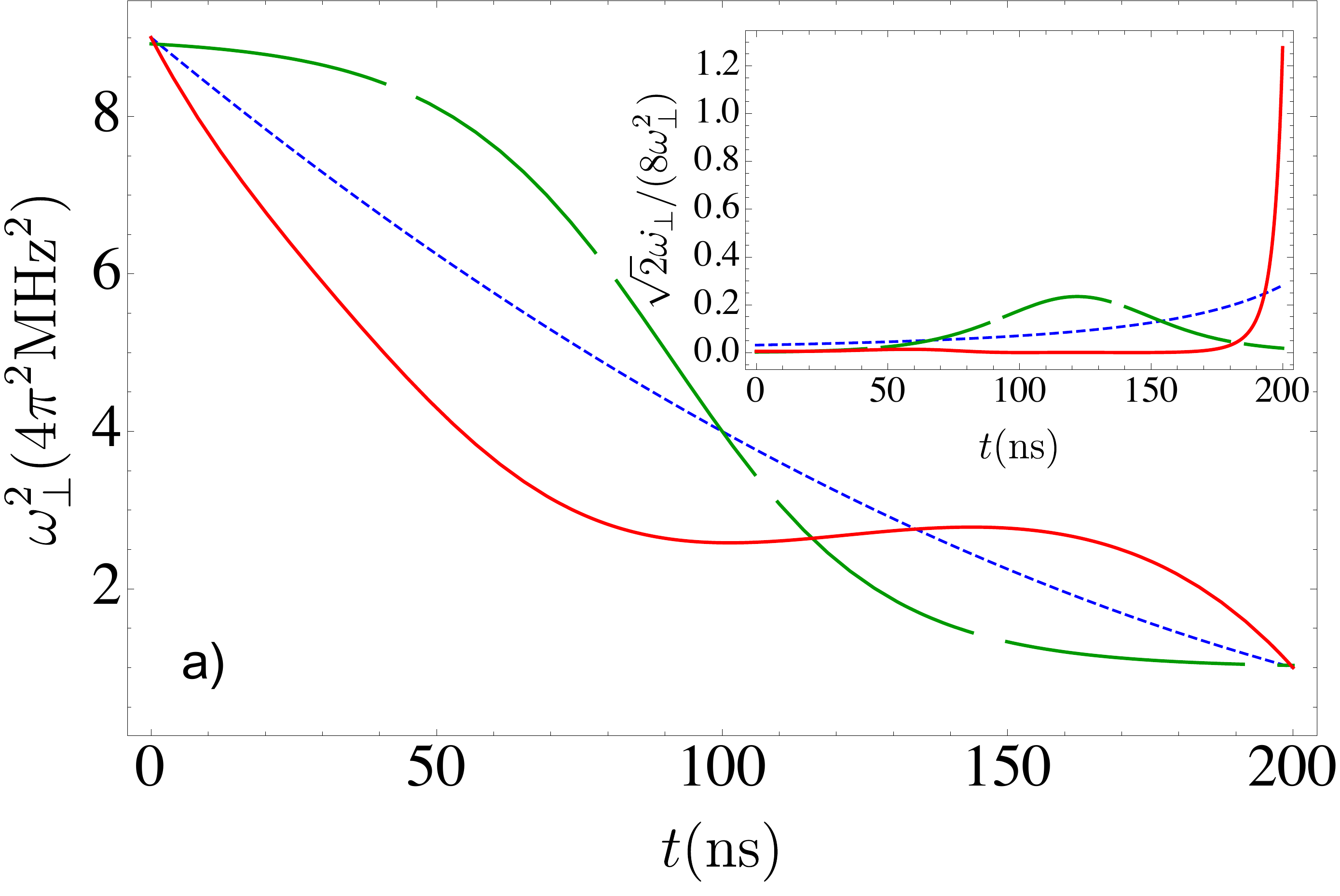}
\includegraphics[width=0.47 \textwidth]{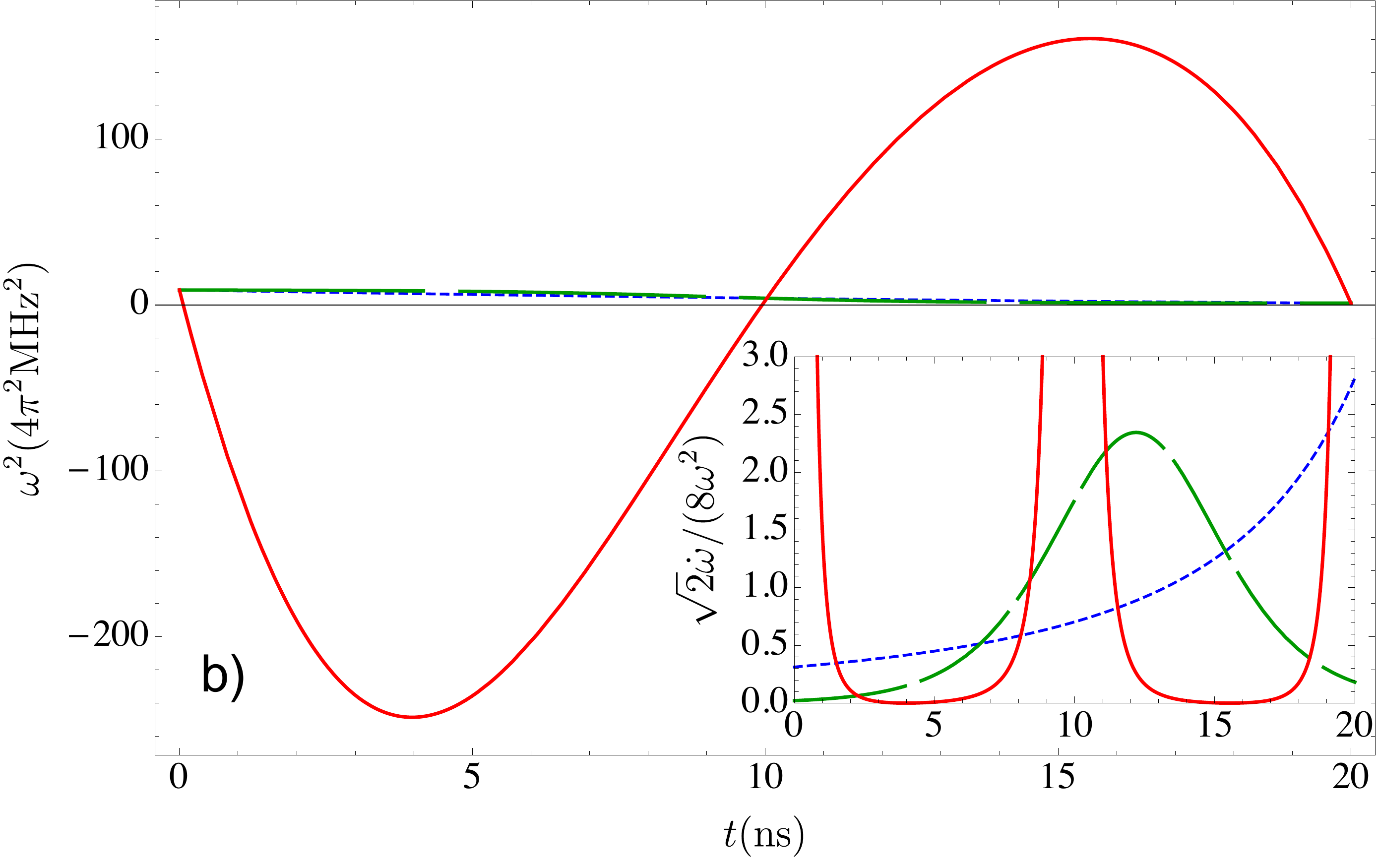}
\caption{Radial confinement squared for a linear ramp (blue short-dashed line), smooth ramp (green long-dashed line) and the shortcut to adiabaticity (red solid line), with (a) parameters chosen to avoid negative $\omega^2$
and (b) allowing negative $\omega^2$ for the shortcut. Here, $\omega_{\mathrm{0}}/(2 \pi)$=3\,MHz and $\omega_{\mathrm{f}}/(2 \pi)$=1\,MHz. Inset: Adiabaticity parameter $\sqrt{2} \dot{\omega}/(8 \omega^2)$ for the three cases. }
\label{fig_shortcuta}
\end{center}
\end{figure}

In the following, we will consider the 3D-Hamiltonian corresponding to
an ion trap symmetrically driven with radiofrequency and end-cap geometry. In order to fulfill Laplace's equation the Hamiltonian describing the trapped ion becomes:
\beq
\label{Hfull}
\hat H(t)=\frac{\hat\textbf{p}^2}{2m}+\frac{m}{2}\omega_z^2(t)\hat z^2+\frac{m}{2}[\Omega(t)+\Delta(t)]^2\hat y^2+\frac{m}{2}[\Omega(t)+\Delta(t))]^2\hat x^2
\eeq
with $\hat\textbf{p}=(\hat p_x,\hat p_y,\hat p_z)$, $\omega_z(t)$ the frequency along the axial $z$-direction, and $\omega_\perp(t)=\omega_x(t)=\omega_y(t)=\Omega(t)+\Delta(t)$ the radial frequencies
produced by the RF and DC voltages in conjunction \footnote{Note that the trapping frequency caused by the pseudopotential and the DC potentials cannot be simply added especially when large voltages are involved (see equation 11 and 15 from \cite{RevModPhys.75.281} for details.)}.
This Hamiltonian has a symmetric radial confinement in the $x$ and $y$-directions that
will be employed as working fluid to produce the heat pump processes. In the following we disregard the effect of control voltages on the longitudinal confinement because the ion is always kept at the extremal point of the longitudinal confinement and we use the longitudinal degrees of freedom as a classical piston being driven. Under this prescription the radial Hamiltonian reads,
\beq
\label{Hz}
\hat H_\perp(t)=\frac{\hat{\mbox{p}}_\perp^2}{2m}+\frac{1}{2}m\omega_\perp^2(t)(\hat y^2+\hat x^2),
\eeq
with $\hat{\mbox{p}}_\perp=(\hat p_x,\hat p_y)$. Defining $\hat{\mbox{r}}_\perp=(\hat x, \hat y)$ we observe that this radial Hamiltonian has the same structure as Eq. (\ref{H1d}), consequently the radial frequency
can be modified from $\omega_\perp(0)=\omega_{\perp,0}$ to $\omega_\perp(t_f)=\omega_{\perp,f}$ through a shortcut
$\omega_\perp^2(t)=\omega_{\perp,0}^2/b^4_\perp-\ddot b_\perp/b_\perp$ with $b_\perp$ satisfying the frictionless boundary conditions (\ref{BC}) with a radial expansion/compression
ratio $\gamma_\perp=(\omega_{\perp,0}/\omega_{\perp,f})^{1/2}$, the same for both the $x$ and $y$ axes.

The shortcut to adiabaticity will be implemented by common voltages on the end-cap electrodes of an ion trap, while the dominant radiofrequency saddle potential has been momentarily turned off.  The differential voltage on the end-caps can be used to control the axial movement of the ion, but can be disregarded here. The radial confinement caused by the radial frequency is only relevant at the turning points of the axial transport, when the ion is coupled to the reservoirs. Alternatively, a linear trap design could be used without a taper, with the radial frequency being switched to different amplitudes in between. The radial trapping potential during the shortcut is applied by a common voltage on the end-cap electrodes, and needs to be matched to the initial and final confinement provided by the pseudopotential. Laplace's equation and the geometric symmetry specifies that $\omega^2$ is inverted with half the magnitude. We have compared three expansion protocols; shortcut, linear and smooth ramp
$\omega(t)=(\omega_0e^{\Gamma t_0}+\omega_fe^{\Gamma t})/(e^{\Gamma t_0}+e^{\Gamma t})$ for the cooling of thermal and coherent states, see Fig. \ref{fig_shortcuta} .

\begin{figure}[t]
\begin{center}
\includegraphics[width=0.46 \textwidth]{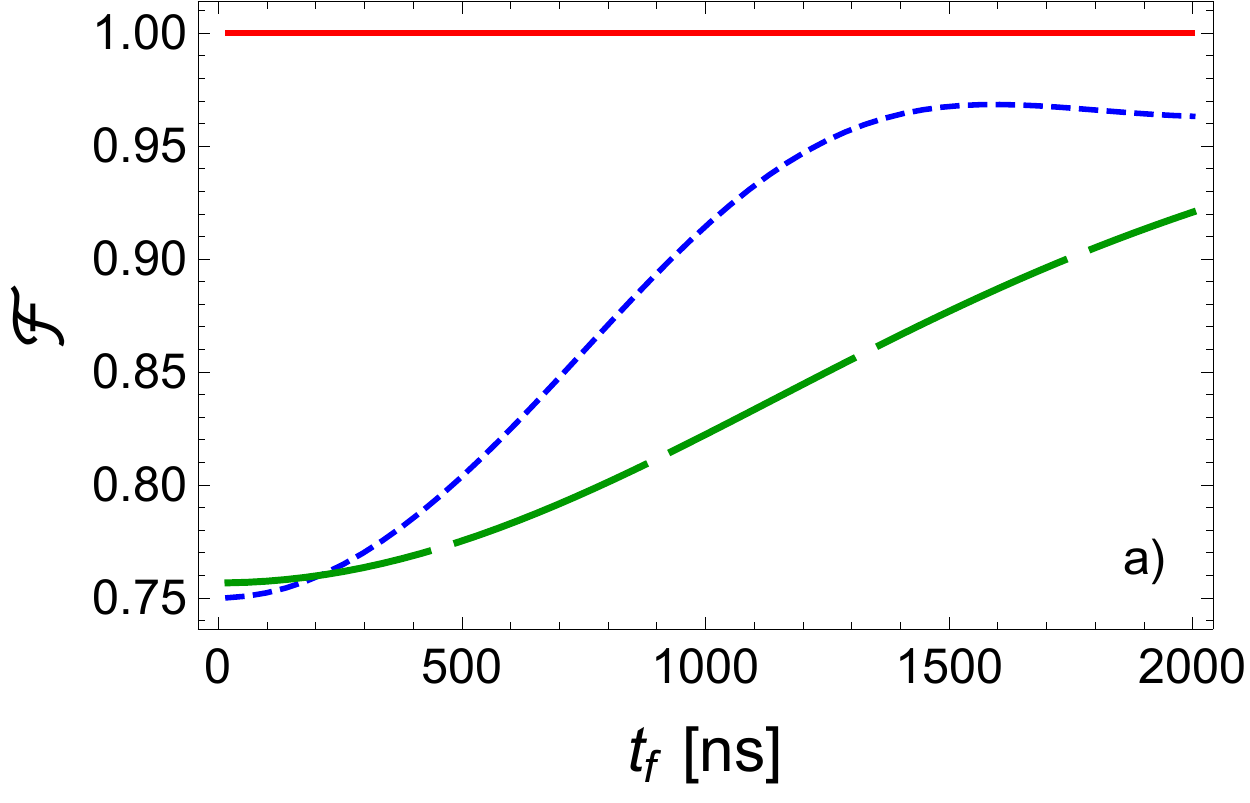}
\includegraphics[width=0.46 \textwidth]{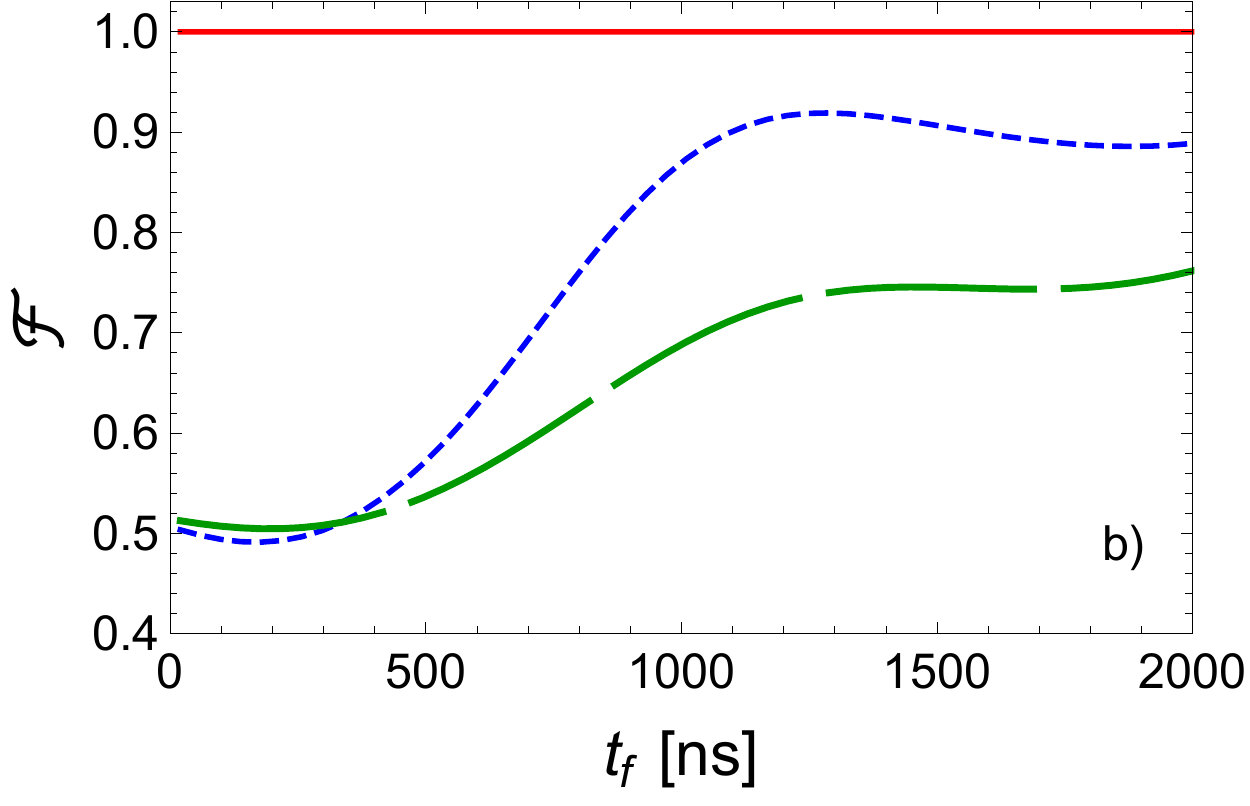}
\caption{Fidelity $\mathcal{F}(\hat\rho(t_f),\hat\rho_{target})$ as a function of the expansion time $t_f$ for three different protocols, shortcut (red solid line), linear ramp (blue short-dashed line), and smooth ramp (green
long-dashed line). In (a) the initial state corresponds to a thermal state with temperature of $2$ mK. In (b) the initial state is a coherent state with a photon number $\alpha_0=1+i$.
Parameters as in Fig. \ref{fig_shortcuta}. }
\label{fig_fidelity}
\end{center}
\end{figure}

The initial thermal state is characterized by the statistical moments $ \bar{\mathbf{X}}_1(0)=\bar{\mathbf{X}}_2(0)=\bar{\mathbf{X}}_5(0)=0$ and
\beq
\bar{\mathbf{X}}_3(0)=l_0^2\coth\bigg(\frac{\beta_0\hbar\omega_{\perp,0}}{2}\bigg), \quad \bar{\mathbf{X}}_4(0)=k_0^2\coth\bigg(\frac{\beta_0\hbar\omega_{\perp,0}}{2}\bigg),
\eeq
with $l_0=\sqrt{\hbar/(2m\omega_{\perp,0})}$ and $k_0=\sqrt{m\hbar\omega_{\perp,0}/2}$
corresponding to a $\hat H(0)$ with a frequency $\omega_\perp(0)=\omega_{\perp,0}$ and inverse temperature $\beta_0$. The target state has similar statistical moments corresponding to a final frequency
$\omega_{\perp,f}$ and inverse temperature $\beta_f=\gamma_{\perp}^2\beta_0$. In Fig. \ref{fig_fidelity}a we plot the fidelity $\mathcal{F}(\hat\rho(t_f),\hat\rho_{target})$ of the evolved state $\hat\rho(t_f)$ compared to
the target thermal state $\hat\rho_{target}$ corresponding to $\hat H(t_f)$ having a frequency $\omega_{\perp,f}$. We observe how the shortcut by construction ensures fidelity one independently of the time employed
to produce the expansion of the harmonic trap whereas  the linear and smooth ramp protocols fail as the process is no longer adiabatic, see insets of Fig. \ref{fig_shortcuta}.

Similarly, we analyze the three previous protocols for the expansion of a coherent state in the trapping potential (\ref{Hfull}). The initial state has the statistical moments
\beqa
\bar{\mathbf{X}}_1(0)&=&2l_0\mbox{Re}(\alpha_0), \quad \bar{\mathbf{X}}_2(0)=2k_0\mbox{Im}(\alpha_0),\\
\bar{\mathbf{X}}_3(0)&=&\bar{\mathbf{X}}_1^2(0)+l_0^2, \quad \bar{\mathbf{X}}_4(0)=\bar{\mathbf{X}}_2^2(0)+k_0^2, \quad \bar{\mathbf{X}}_5(0)=4\hbar\mbox{Re}(\alpha_0)\mbox{Im}(\alpha_0)\nonumber
\eeqa
associated with $\hat H(0)$ and $\omega_{\perp,0}$. At $\hat H(t_f)$ the target state has similar statistical moments with $\omega_{\perp}(t_f)=\omega_{\perp,f}$ and
photon number $\alpha_f=\alpha_0 e^{-ig\omega_{\perp,0}}$ with $g=\int_0^{t_f}dt'/\rho^2$. As for the case of thermal states we observe in Fig. \ref{fig_fidelity}b how the shortcut drives the initial system
until the desired target state independently of the expansion time $t_f$.

\begin{figure}[t]
\begin{center}
\includegraphics[width=0.7 \textwidth]{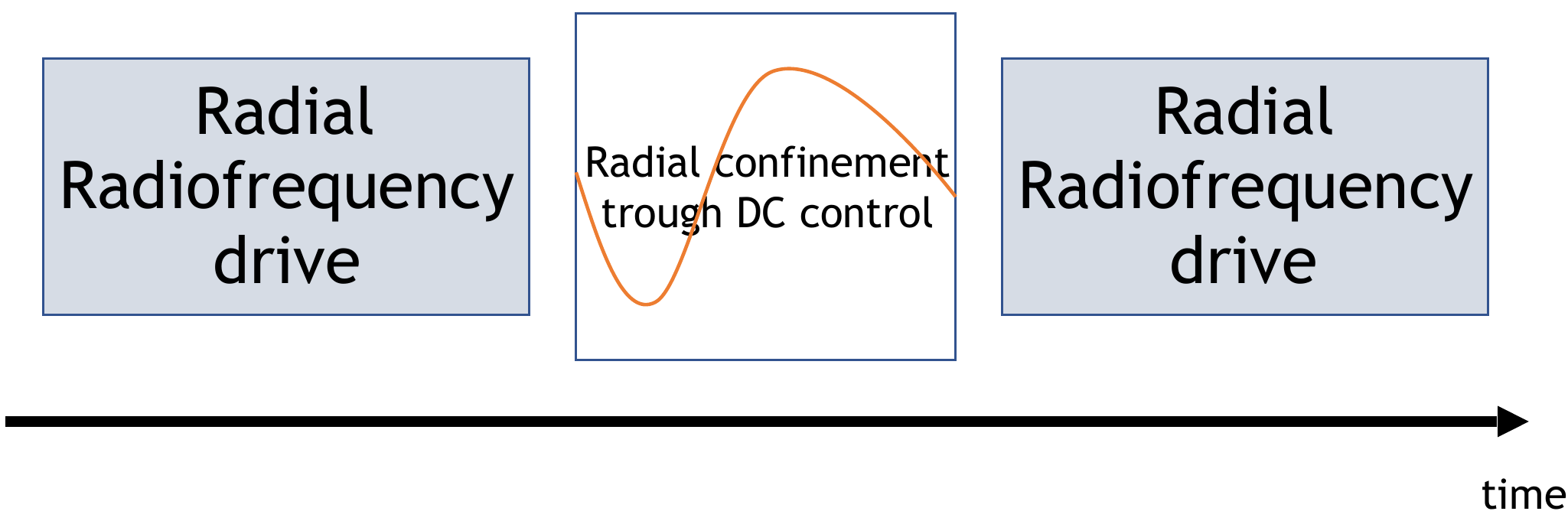}

\caption{Experimental control sequence. The radial radiofrequency drive is switched off during the application of the shortcut to adiabaticity protocol on the dc electrodes, the shortcut changes the radial confinement.}
\label{fig_timing}
\end{center}
\end{figure}

\begin{figure}[t]
\begin{center}
\includegraphics[width=0.8 \textwidth]{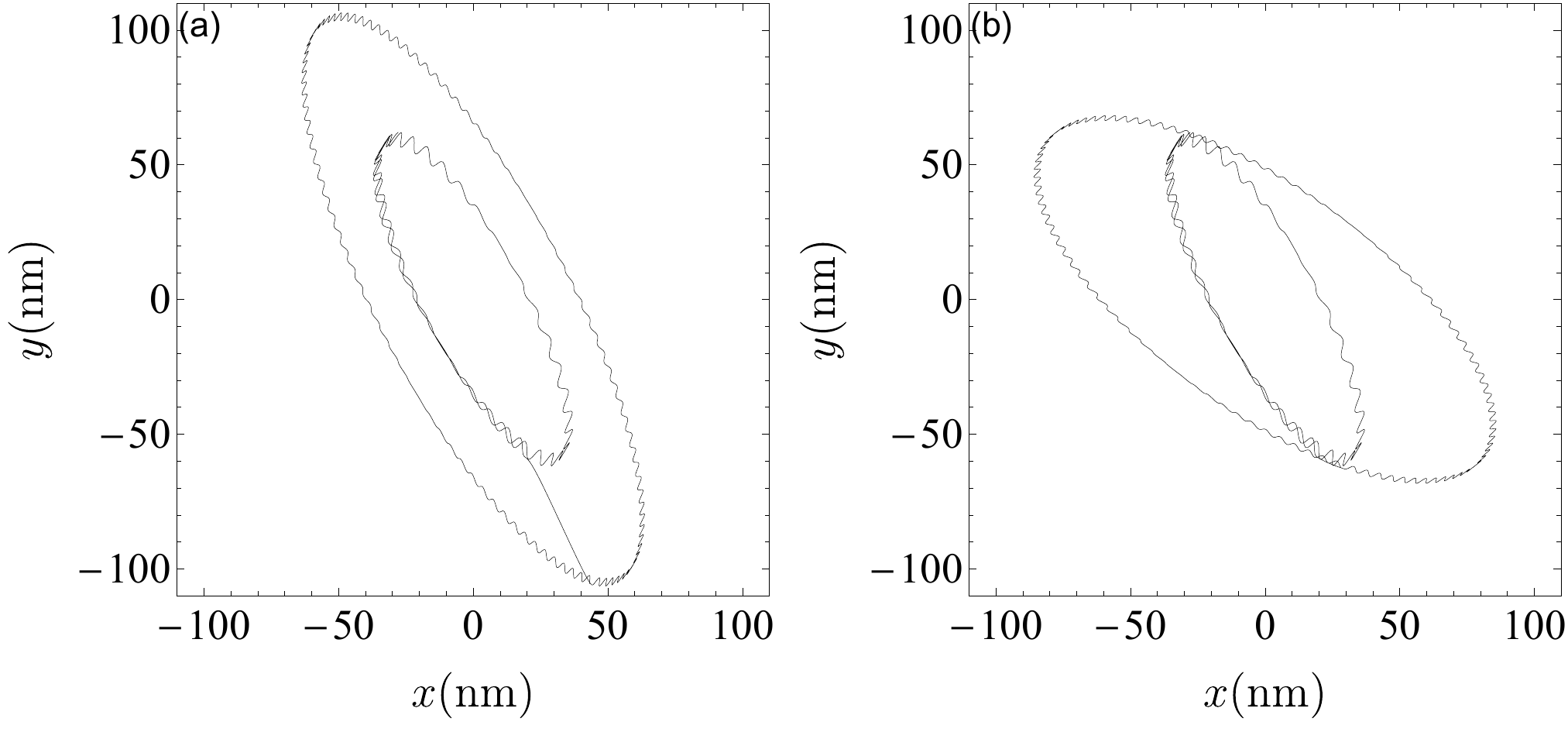}
\caption{Simulation of the classical radial trajectory of the ion including the micro-motion effect for the whole experimental control sequence depicted in Fig. \ref{fig_timing}. The expansion time is $20$ ns for both the shortcut (a) and the equivalent linear ramp (b). In both cases the initial radial trapping frequencies are $\omega_\perp(0)/(2\pi)=3$MHz and $\omega_\perp(\textrm{20ns})/(2\pi)=1$MHz. The radiofrequency drive was set to $\omega_\textrm{RF}/(2 \pi)=100$MHz with the axial trapping
frequency being $\omega_z/(2\pi)=100$kHz. Rest of parameters: $m=40$ a.m.u.}
\label{fig_trajectory}
\end{center}
\end{figure}

Figure \ref{fig_timing} shows the whole control sequence responsible for the shortcut to adiabaticity protocol which includes antitrapping potentials for short compression cycles. The radiofrequency is switched off during that time such that the DC control potentials can be kept at lower voltages. By construction the protocol keeps the fidelity at 1, but stable trapping conditions have to be maintained due to the anti-trapping potentials involved. In Fig.~\ref{fig_trajectory} we have verified that indeed phase stable trapping can be maintained due to the shortness of the anti-trapping potentials. We have included in the dynamics the whole experimental control sequence Fig. \ref{fig_timing}, where the trapping potential is given by Eq. (\ref{Hfull}) and the micromotion exerted on the ion due to the rf-driving has been taken into account. To include this micromotion, a simulation based on the velocity Verlet method was performed.
Both the radiofrequency drive $\omega_{RF}/(2 \pi)$ and the axial trapping $\omega_z/(2\pi)$
frequencies were set to $100$ kHz. In order to avoid instability due to micromotion, the corresponding radiofrequency period is shorter than the shortcut duration produced in $20$ ns.
For this expansion time (see Fig.~\ref{fig_shortcuta}b), the adiabaticity parameter goes beyond the adiabatic regime for the linear and smooth ramps, thus making the shortcut necessary to ensure a perfect driving. Note, due to the zero-crossings of $\omega^2$, the adiabaticity parameter diverges, but this does not compromise the effectiveness of the shortcut. This is also apparent in Fig.~\ref{fig_trajectory}a, where one can observe that a phase relation is maintained before and after the shortcut. In contrast, in Fig. ~\ref{fig_trajectory}b, although the ion remains
trapped after the linear ramp the final evolved state is excited. The excitations modify the ion oscillations rotating the axis of the ellipse with respect to the original direction that corresponds to the final unexcited state.
\section{Discussion}
 Making use of shortcuts to adiabaticity we have improved the efficiency of a heat pump for a single ion. The expansion protocol allows ultra-fast and high-fidelity processes through the use of transient non-confining potentials.
The stability of the potential has been analyzed and the experimental feasibility discussed. The shortcut control has been improved according to experimental constrains, in particular minimizing the required power and
thus reducing the effect of noise produced by the controls.
These improved controls could be useful since
efficient heat pump extraction protocols provide new cooling mechanisms and constitute the basis of  stroke heat engines/refrigerators \cite{Kosloff1984} allowing us to test the laws of thermodynamics and get closer to the absolute zero temperature \cite{Torrontegui2013} in the single particle domain.
The possibility to design different refrigerators based on the Otto cycle according to the performance of each stroke offers a new venue to design new heat pump protocols. As example, not only optimizing the compression/expansion strokes but also designing efficient trapping
potentials at the isochores for the thermalization processes by controlling the trap frequencies $\omega(t)$. Additionally, using the temperature of the bath as a control could lead to new shortcut to adiabaticity such that the optimal performance of the heat pump would be achieved. These extensions are of additional interest also to different refrigerators types like the continuous refrigerator where the ion is in continuous contact
with the bath \cite{Kosloff2014}, which might be easier to implement experimentally.
\ack
We acknowledge funding from MINECO/FEDER (Grants No. FIS2015-70856-P and No. FIS2015-67161-P), CAM PRICYT project QUITEMAD+CM S2013-ICE2801, and Basque Government (Grant No. IT986-16).
K. S. acknowledges support by the german science foundation through the single ion heat engine project.
\\
\\
\appendix
\section{Invariant-based inverse engineering for pure states}
\label{LR_pure}
Related to any Hamiltonian $\hat H(t)$ there are invariants of motion \cite{Lewis1969}
\beq
\label{dynI_ap}
i\hbar\frac{\partial\hat I(t)}{\partial t}-[\hat H(t),\hat I(t))]=0,
\eeq
with constant expectation values for any wave function satisfying the time-dependent Schr\"odinger equation
\beq
i\hbar\frac{\partial}{\partial t}|\Psi(t)\rangle=\hat H(t)| \Psi(t)\rangle.
\eeq
The invariant expands  an orthonormal basis $|\phi_n(t)\rangle$ with constant eigenvalues $\lambda_n$,
\beq
\label{inv}
\hat I(t)=\sum_n|\phi_n(t)\rangle\lambda_n\langle\phi_n(t)|.
\eeq
These states can be used to express the dynamical wave function as a linear superposition of the ``dynamical modes"
\beq
\label{wf}
|\Psi(t)\rangle=\sum_n c_n|\psi_n(t)\rangle \quad\mbox{with} \quad |\psi_n(t)\rangle=e^{i\alpha_n(t)}|\phi_n(t)\rangle,
\eeq
 $c_n$ being the constant time-independent coefficients of the expansion with the Lewis-Riesenfeld phases defined as \cite{Lewis1969}
\beq
\label{phaseLR}
\alpha_n(t)=\frac{1}{\hbar}\int_0^t dt^\prime\langle \phi_n(t^\prime)\bigg|i\hbar\frac{\partial}{\partial t^\prime}-\hat H(t^\prime)\bigg|\phi_n(t^\prime)\rangle.
\eeq
Suppose that we want to drive the system by changing a control parameter $\epsilon(t)$ from an initial Hamiltonian $\hat H(\epsilon(t=0))$ with $\epsilon(t=0)=\epsilon_0$ to a final configuration governed by $\hat H(\epsilon(t=t_f))$, where $\epsilon(t=t_f)=\epsilon_f$
in such a way that the populations in the initial and final instantaneous basis are the same but
transitions at intermediate times are allowed\footnote[1]{More controls $\epsilon_1(t), \dots \epsilon_n(t)$ can be considered but a singe control is assumed for simplicity.}.
Our aim is to deduce the time dependency of the control $\epsilon(t)$ that enables us to perform this task. We assume that the structure of the Hamiltonian controlling the dynamics of the system is known, i.e., the dependency of $\hat H=\hat H(\epsilon)$ as a function of $\epsilon$
is known but not the time dependency of $\epsilon=\epsilon(t)$, which is our target. Once $\hat H(\epsilon)$ is known, a related invariant can be found using Eq. (\ref{dynI_ap}) and subsequently its eigenvectors $|\phi_n(\epsilon)\rangle$\footnote[2]{The relative phases between the eigenstates of the invariant allow different definitions of the $|\phi_n\rangle$ states; consequently the Lewis-Riesenfeld phase (\ref{phaseLR}) is non-unique.} and eigenvalues deduced. Then the state of the system at any time will be described
by Eqs. (\ref{wf}) and (\ref{phaseLR}) evolving during the whole process as a linear combination of the dynamical modes. Generally, notice that $\hat I(t=0)$ does not commute with $\hat H(t=0)$, then the eigenstates of the invariant do not coincide with those of the Hamiltonian.
A similar situation occurs at $t=t_f$. Imposing the frictionless conditions $[\hat I(0),\hat H(0)]=0$ and $[\hat I(t_f),\hat H(t_f)]=0$ will allow us to deduce a control strategy $\epsilon=\epsilon(t)$ that guarantees a perfect state evolution without
final excitations such that the initial and final states are compatible with the initial/final Hamiltonians \cite{Chen2010_063002, Torrontegui2013_117}.
\section{Fast expansion and compression of a harmonic trap}
\label{reverse}
In this section we will apply the general formalism to a particular case corresponding to the expansion/compression of a time-dependent harmonic
potential \cite{Chen2010_063002, Torrontegui2013_117, del2014more, Torrontegui2012_033605, Julia-Diaz2012, Yuste2013, Levy2017, Levy2018}.
We consider a particle of mass $m$ trapped by an effectively 1D time dependent harmonic potential
\beq
\label{H1d_ap}
\hat H(t)=\frac{\hat p^2}{2m}+\frac{1}{2}m\omega^2(t)\hat q^2
\eeq
with an initial frequency $\omega(0)=\omega_0$ and a final trapping configuration that corresponds to $\omega(t_f)=w_f$. For $\omega_0>\omega_f$ ($\omega_0<\omega_f$) the process corresponds to an expansion (compression) of the
trap. Our goal is to find the control $\omega(t)$ so that the system evolves from any eigenstate $|n(0)\rangle$ of $\hat H(\omega_0)$ at $t=0$ to the corresponding eigenstate $|n(t_f\rangle)$
of $\hat H(\omega_f)$ at $t=t_f$. A dynamical invariant of the Hamiltonian (\ref{H1d_ap}) reads \cite{Lewis1982}
\beq
\label{iho_ap}
\hat I(t)=\frac{1}{2m}[b(t)p-m\dot b(t)q]^2+\frac{1}{2}mc^2\frac{q^2}{b^2(t)},
\eeq
where $b(t)$ is a free function of time satisfying the Ermakov equation \cite{Ermakov1880}
\beq
\label{erm_ap}
\ddot b(t)+\omega^2(t)b(t)=\frac{c^2}{b^3(t)},
\eeq
where for convenience we set the constant $c=\omega_0$. Defining $\hat\pi=b\hat p-m\dot b\hat q$ which is the conjugate momentum of $\hat qb$, we notice that the invariant (\ref{iho_ap}) has the structure of a harmonic oscillator with constant
frequency $c=\omega_0$. After computing the phases $\alpha_n(t)=-(n+1/2)\omega_0\int_{0}^tdt^{\prime}/b^2(t^{\prime})$ and using Eq. (\ref{wf}) we found the wave function of the system at any time. Considering a single mode with $\omega_0^2>0$
\beqa
\label{modeHO}
\Psi_n(q,t)\equiv\langle\hat q|\Psi_n(t)\rangle&=&\bigg(\frac{m\omega_0}{\pi\hbar}\bigg)^{1/4}\frac{e^{i(m/2\hbar)(\dot b/b+i\omega_0/b^2)q^2}}{(2^nn!b)^{1/2}} \nonumber\\
&\times&e^{-i(n+1/2)\omega_0\int_0^tdt^{\prime}/b^2}H_n\bigg[\sqrt{\frac{m\omega_0}{\hbar}}\frac{q}{b}\bigg],
\eeqa
with $H_n$ the $n$-order Hermite polynomial. The average energy for this state becomes \cite{Chen2010_063002}
\beq
\langle\hat H(t) \rangle_n=\frac{(2n+1)\hbar}{4\omega_0}\bigg(\dot b^2(t)+\omega^2(t)b^2(t)+\frac{\omega_0^2}{b^2(t)}\bigg),
\eeq
having a zero average position, a standard deviation
\beq
\Delta q_n^2(t)=\int_{-\infty}^{\infty} dq q^2|\Psi_n(q,t)|^2=\hbar b^2(t)\bigg(\frac{n+1/2}{m\omega_0}\bigg),
\eeq
and gives a physical meaning to $b(t)$.
To set $|\Psi(0)\rangle$ and $|\Psi(t_f)\rangle$ as eigenstates of the initial and final Hamiltonians we impose the frictionless conditions $[\hat H (t_b)=\hat I(t_b)]=0$ at the boundary times $t_b=0, t_f$ that
implies $b(0)=1, b(t_f)=\gamma=(\omega_0/\omega_f)^{1/2}$, and $\dot b(0)=\dot b(t_f)=\ddot b(0)=\ddot b(t_f)=0$.
These boundary conditions are easily obtained making $\hat I(0)=\hat H(0)$ and $\hat I(t_f)=\gamma\hat H(t_f)$.
The conditions for the second derivative follow from Eq. (\ref{erm_ap}) that holds at all time in order to impose $\hat I(t)$ as a dynamical invariant of $\hat H(t)$. Then any $b(t)$ fulfilling the previous six conditions
at the extremes will produce the desired driving
\beq
\label{w_sta}
\omega^2(t)=\frac{\omega_0^2}{b^4(t)}-\frac{\ddot b(t)}{b(t)}
\eeq
between the states of $\hat H(0)$ and $\hat H(t_f)$ independently of the expansion/compression time $t_f$. 
In order to satisfy (\ref{BC}) we interpolate $b(t)=\sum_{i=0}^5a_it^i$ with at least the same number of coefficients $a_i$ as conditions over $b.$ Solving for the coefficients we find $b(t)=6(\gamma-1)s^5-15(\gamma-1)s^4+10(\gamma-1)s^3+1$
where $s:=t/t_f$.
We can take advantage of the non-uniqueness of $b(t)$ at intermediate times to design more sophisticated $b(t)$ functions and additionally minimize or impose possible
experimental constraints \cite{Levy2017, Levy2018, Stefanatos2010, Chen2011_043415, Torrontegui2017}.
\\

\bibliographystyle{iopart-num}
\bibliography{heat}

\end{document}